# Metallization of hydrogen by intercalating ammonium ions in metal fcc lattices at low pressure

Zhongyu Wan[1], Ruiqin Zhang[1][2]*

[1] Department of Physics, City University of Hong Kong, Hong Kong SAR 999077, People's Republic of China

[2] Beijing Computational Science Research Center, Beijing, 100193, People's Republic of China

*E-mail: aprqz@cityu.edu.hk

**Abstract**

Metallic hydrogen is capable of showing room temperature superconductivity, but its experimental syntheses are extremely hard. Therefore, it is desirable to reduce the synthesis pressure of metallic hydrogen by adding other chemical elements. However, for most hydrides, the metallization of hydrogen by "chemical precompression" to achieve high-temperature superconductivity still requires relatively high pressure, making experimental synthesis difficult. How to achieve high-temperature superconductivity in the low-pressure range ($\leqslant$ 50 GPa) is a key issue for realizing practical applications of superconducting materials. Toward this end, this work proposes a strategy for inserting ammonium ions in the fcc crystal of metals. High-throughput calculations of the periodic table reveal 12 elements which can form kinetically stable and superconducting hydrides at low pressures, where the highest superconducting transition temperatures of $AlN_2H_8$, $MgN_2H_8$ and $GaN_2H_8$ can reach up to 118.40, 105.09 and 104.39 K. Pressure-induced bond length changes and charge transfer reveal the physical mechanism of high-temperature superconductivity, where the H atom continuously gains electrons leading to the transition of $H^+$ ions to atomic H, facilitating metallization of hydrogen under mild pressure. Our results also reveal two strong linear scalar relationships, one is the H-atom charge versus superconducting transition temperature and the other is the first ionization energy versus the highest superconducting transition temperature. Besides, $ZnN_2H_8$, $CdN_2H_8$, and $HgN_2H_8$ were found to be superconductors at ambient pressure, and the presence of interstitial electrons suggests that they are also electrides, whose relatively low work functions (3.03, 2.78, and 3.05 eV) imply that they can be used as catalysts for nitrogen reduction reactions as well.

Superconducting materials attract the wide attention of scientists because of their zero resistance and anti-magnetic properties[1] in their superconducting states[2] at a temperature below their superconducting transition temperatures ($T_c$). Therefore, the $T_c$ is the key to determining whether the material can be used in practice.[3] However, most superconductors do not have high $T_c$,[4] which limits the wide application of superconducting materials and makes the search for high-temperature superconductors become one of the most critical issue in condensed matter physics. Ashcroft et al.[5] predicted that the monomeric H would transform to the metallic state at high pressure with $T_c$ reaching room temperature, and metallic hydrogen was experimentally observed at the pressure of 495 GPa.[6]

However, the pressure required to prepare metallic hydrogen experimentally is too high. Despite the high $T_c$, the high pressure required is still unfriendly to experimentalists. So, it is necessary to design high-temperature superconductors that can be easily synthesized.[7] In recent years, more and more theoretical work has shown that "chemical precompression"[8] realized by adding other elements to the hydrogen monomers allows the preparation of metallic hydrogen with much lower pressure.[9] Examples include the high-temperature superconductors $H_3S$,[10] $SiH_4$,[11] $CaH_6$,[12] $YH_6$,[13] and $LaH_{10}$.[14] However, the synthesis pressure of these hydrogen-rich compounds still has to be more than 100 GPa, meaning that they can only be synthesized in the micron scale in expensive diamond anvil cell,[15] which is still uneasy for experimental synthesis.

Fortunately, ternary hydrides, such as $LaBH_8$,[16] $LaBH_9$,[17] and $LaBeH_8$,[18] exhibit stronger "precompression" effects. These compounds are predicted to be potential superconductors and achieve high-temperature superconducting states at relatively low pressures. This implies that they have the potential to be synthesized as millimeter-scale products by inexpensive large volume cubic pressure cells,[19] greatly enhancing the potential for industrial applications. Recently Xiang et al.[20] predicted that $KB_2H_8$ obtained by introducing $[BH_4]$ units into the fcc lattice of potassium could have a $T_c$ of 146 K at 12 GPa. In addition, Zhong et al.[21] inserted methane into the fcc lattice of Na, K, Mg, Al, and Ga, and theoretically predicted a series of excellent BCS superconductors, such as $MgC_2H_8$ ($T_c$ = 55 K, 40 GPa) and $AlC_2H_8$ ($T_c$ = 67 K, 80 GPa), which provided us with a reference for designing high-temperature superconductors under mild conditions. Element N has similar electronegativity and atomic structure compared to B and C. The difference is that the 2p orbital of the N atom can only accept 3 electrons, which means the H atom in $[NH_4]$ cannot form a complete $H^+$ ion and has a certain degree to turn into atomic H, which is favorable to turn into a high-temperature superconductor as atomic metallic hydrogen.[22]

Therefore, in this work, ammonium ions are attempted to be intercalated inside the fcc lattice of a series of representative metals to realize chemical precompression of H with the help of N and the other element, making it possible to metallize H at lower pressures. Our study is expected to provide theoretical guidance for the experimental synthesis of high-temperature superconductors under mild conditions.

The Vienna Ab-initio Simulation Package [23] is used to perform density functional calculations, and the projected-augmented wave pseudopotentials method[24] is employed to describe the electron-ion interaction. Generalized gradient approximation and Perdew-burke-ernzerhof as exchange-correlated functional[25] with a cutoff energy of 900 eV and k-point grid density of $0.03 \times 2\pi$ based on the Monkhorst-Pack method[26]. The relaxed ionic positions and cell parameters provided energy and forces smaller than $10^{-5}$ eV and 0.002 eV/Å, respectively. The QUANTUM ESPRESSO package [27] is used to calculate the electron-phonon coupling (EPC) properties in linear response theory, the ultra-soft pseudopotential with a kinetic cutoff energy of 160 Ry and a charge density cutoff energy of 1500 Ry, respectively. Self-consistent field (SCF) calculations are performed on a fine k-point grid with a density of $40 \times 40 \times 40$, the Methfessel-Paxton first-order spread is set to 0.01 Ry, and an irreducible q-point grid with a phonon density of $10 \times 10 \times 10$ is used to further calculate the phonon property. The Coulomb pseudopotential parameter $\mu^*= 0.10$ is used in the modified McMillan-Allen-Dynes equation[28] to estimate the superconducting transition temperature.

The kinetic stability determined by phonon vibrations is a prerequisite for the compound's superconductivity. Typical pressure points of 10, 30, and 50 GPa were used to test the kinetic stability of the compound $XN_2H_8$. High-throughput calculations reveal that the fcc lattices of 12 elements remain stable after being intercalated by $NH_4$. It is noteworthy that these elements have similar electronegativity (from 1.2 to 1.9) and valence electron numbers (1 or 2). Having a slightly lower electronegativity than H (2.1) allows the charges on their atoms to migrate to the H atoms, avoiding the electrons being bound to the localized regions of the atoms, leading to the reduced stability caused by the Jahn-Teller effect.[29] The Pauling electronegativity 2.4 of Au is responsible for the instability of Au in I B group together with Cu and Ag. In contrast, Ca, Sr, and Ba, which are in II A group with Be and Mg, are more likely to provide electrons to H due to their smaller electronegativity, leading to the instability caused by the localization of electrons on the H atom. The crystal structure of the 12 $XN_2H_8$ can be seen in Table S1.

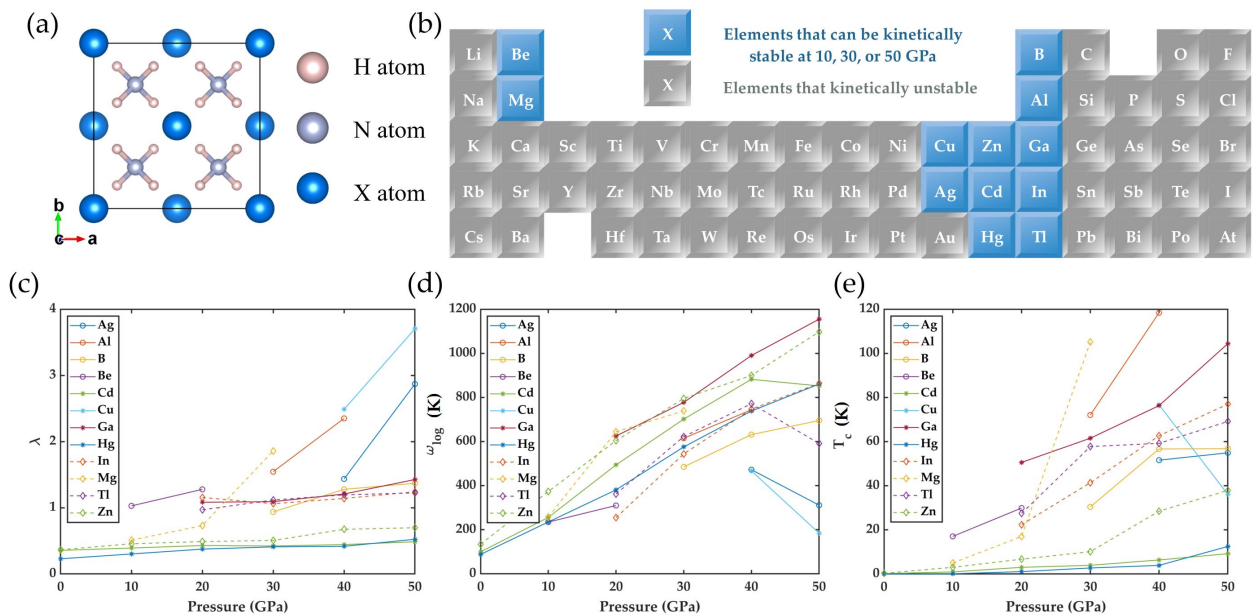

**Figure 1.** (a) The crystal structure of $XN_2H_8$. (b) Kinetic stability at 10, 30, and 50 GPa. (c) Pressure-induced electron-phonon coupling parameters $\lambda$, (d) log-averaged phonon frequency $\omega_{\log}$, and (e) superconducting transition temperature $T_c$.

These 12 $XN_2H_8$ compounds were further detailed to calculate their electron-phonon coupling properties from 0 to 50 GPa. Figures 1(c)-(e) demonstrate the variation of $\lambda$, $\omega_{\log}$ and $T_c$ with pressure, and surprisingly, three ternary hydrides exhibit high temperature superconductivity ($T_c$ >77K), which are $AlN_2H_8$ (40 GPa, $\lambda$=2.35, $\omega_{\log}$=743.95 K, $T_c$=118.40 K), $MgN_2H_8$ (30 GPa, $\lambda$= 1.86, $\omega_{\log}$=738.87 K, $T_c$=105.09 K), and $GaN_2H_8$ (50 GPa, $\lambda$=1.43, $\omega_{\log}$=1154.92 K, $T_c$=104.39 K). For X=Ag ($\lambda$=2.87, $\omega_{\log}$=311.33 K, $T_c$=54.92 K) and Cu (40GPa, $\lambda$=2.49, $\omega_{\log}$=466.69 K, $T_c$=76.65 K). Although they have higher $\lambda$ (Figure 1(c)), the low $\omega_{\log}$ (Figure 1(d)) causes the $T_c$ to not reach above 77 K. The strong EPC and the high $\omega_{\log}$ may be the reasons for the high $T_c$.

Pressure not only changes physical properties but also affects the structure of the crystal. Usually, the pressure makes the lattice more compact and leads to a decrease in the atomic spacing. On the contrary, with increasing pressure, the N-H bond length in [$NH_4$] increases in this system, probably due to the influence of H by the neighboring X atoms.

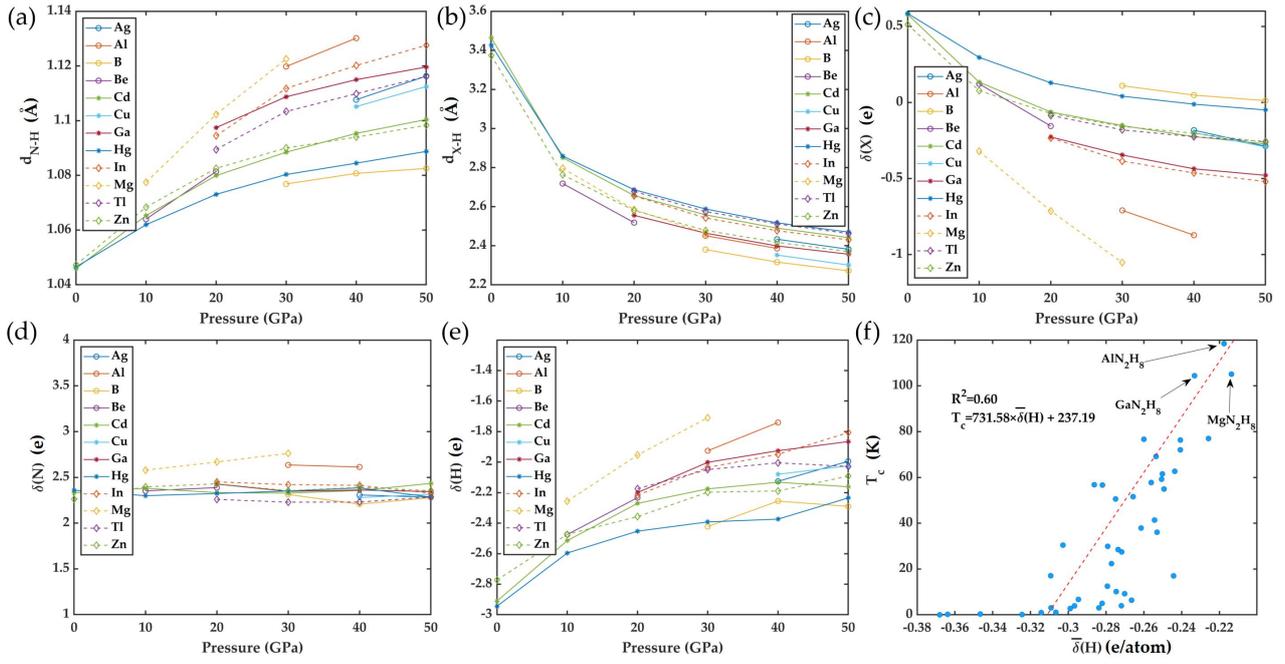

**Figure 2.** Effect of pressure on (a) N-H distance, (b) X-H distance in the lattice. Pressure-induced changes in the total Bader charge of (c) X , (d) N , and (e) H atoms. And (f) scatter plot of $T_c$ vs. average Bader charge per H atoms based on 48 data points.

Figure 2(b) reveals the distances between X and H atoms at high pressure. The increasing pressure makes their distances shorter, and the electron cloud with a higher degree of overlap makes their interactions stronger. The attractive force originating from the X atoms is the possible reason for the shorter N-H bond length. Bader charge [30] is performed to obtain the charge transfer affected by the bond length change. Figures 2(c)-(e) show the total Bader charge for each type of atom in $XN_2H_8$. Bader charge of N varies very little, so the pressure does not make the charge on N migrate significantly. Compared to the N atom, the tremendous pressure causes a significant decrease in the number of electrons on X and a significant increase in the number of electrons on H. This indicates that the pressure causes charge transfer between the X and H atoms, which explains the shortening of the X-H distance. Furthermore, our results reveal the trend of atomization of H in the $XN_2H_8$ system, where the high electronegativity of N makes it attract the 1s electrons of H to fill the N-2p orbitals in the [$NH_4$] unit. However, the X atom tends to provide electrons for H due to its lower electronegativity, which makes the [$NH_4$] gain more electrons and become unstable, so as to make it easier for the N-H bond to dissociate and the $H^+$ ion to become atomic H and reach a similar environment as the H atom in metallic hydrogen. The atomization of $H^+$ gives electrons at lower levels the opportunity to reach the Fermi level, which may enhance the superconductivity. This conclusion is further verified in Figure 2(f), where the average Bader charges per H ($\bar{\delta}$) are -0.218,

-0.214, and -0.233e in three high-temperature superconductors AlN$_2$H$_8$, MgN$_2$H$_8$, and GaN$_2$H$_8$, respectively, which are closer to 0 compared to the low $T_c$ ones. $T_c$ has a linear scalar relationship with $\bar{\delta}(H)$. The stronger the tendency of H$^+$ ions to transform into atomic H, the higher $T_c$ of this superconductor.

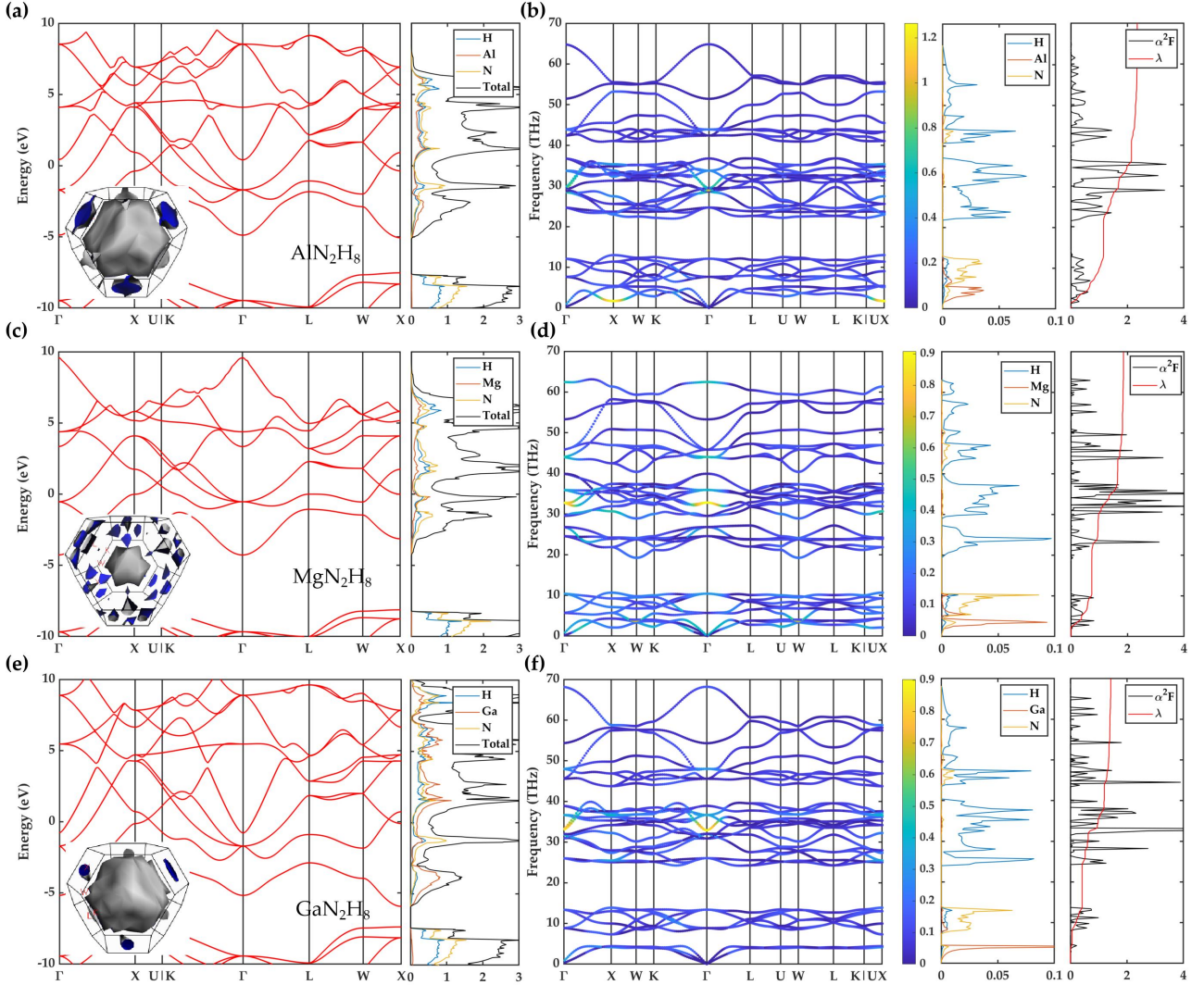

**Figure 3.** (a) and (b) are band structure, the density of states, phonon dispersion curves, phonon density of states (PHDOS), and Eliashberg spectral function of AlN$_2$H$_8$ at 40 GPa. (c) and (d) are those of MgN$_2$H$_8$ at 30 GPa. (e) and (f) are those of GaN$_2$H$_8$ at 50 GPa, where the colors in the phonon dispersion curves are weights on $\lambda$.

Considering the charge transfer between X and H, the properties of X may also affect the superconductivity. Figure S1 shows that the highest $T_c$ of XN$_2$H$_8$-type compounds is strongly correlated with the first ionization energy of X atoms, because the lower the first ionization energy of X atoms because the lower the first ionization energy, the more accessible for the electrons on X to break away from the nucleus, and the easier for charge transfer to happen between X and H. In addition to charge transfer, its electronic structures will determine outstanding physical properties.

Therefore, it is necessary to study the electronic structures of superconductors with high $T_c$ to gain physical insight.

Despite having different types of X atom and external pressures, their band structures show similar characteristics (Figure 3(a), (c), and (e)), which provide theoretical mechanisms for the design of high $T_c$ superconducting materials. The electronic states of N and H at higher energy levels are much less than those at lower levels (<-7 eV), and N and H located at lower levels have the same trend of DOS, which confirms the existence of N-H bonds in the system. The DOS of X, N, and H atoms above 7 eV has a similar trend, indicating a strong coupling between these three types of atoms, consistent with the characteristics of high-temperature superconductors.[31] In addition, the DOS of the X atom appears at the >-7eV level instead of the <-7eV one because there is some charge transfer and interaction between X and H, but no stable chemical bond is formed. Although the differences in pressure and elements lead to inconsistencies in Fermi surfaces, the Fermi surface near the central Γ-point has a similar spherical shape, which has the same behavior as the Fermi surfaces of $AlC_2H_8$ and $MgC_2H_8$.[21]

The phonon dispersion curves (Figure 3(b), (d), and (f)) reveal the distribution of phonons at different frequencies. From the PHDOS, the low-frequency phonons below 15 THz mainly originate from X and N atoms, where the phonons of X atoms are lower than those of N due to the relatively larger atomic masses of Al, Mg, and Ga. The lower atomic mass of the H atom allows it to dominate the PHDOS for high-frequency phonons above 15 THz, and the contributions of high-frequency phonons above 15 THz to $\lambda$ are 51%, 60%, and 71%, respectively, indicating that the H atom dominates the electron-phonon coupling. Similarly, the phonons around 30 THz at the Γ point contribute significantly to the EPC. In addition, phonon softening is observed near the X point on $AlN_2H_8$, resulting in $AlN_2H_8$ having a slightly higher $T_c$ than $MgN_2H_8$ and $GaN_2H_8$.

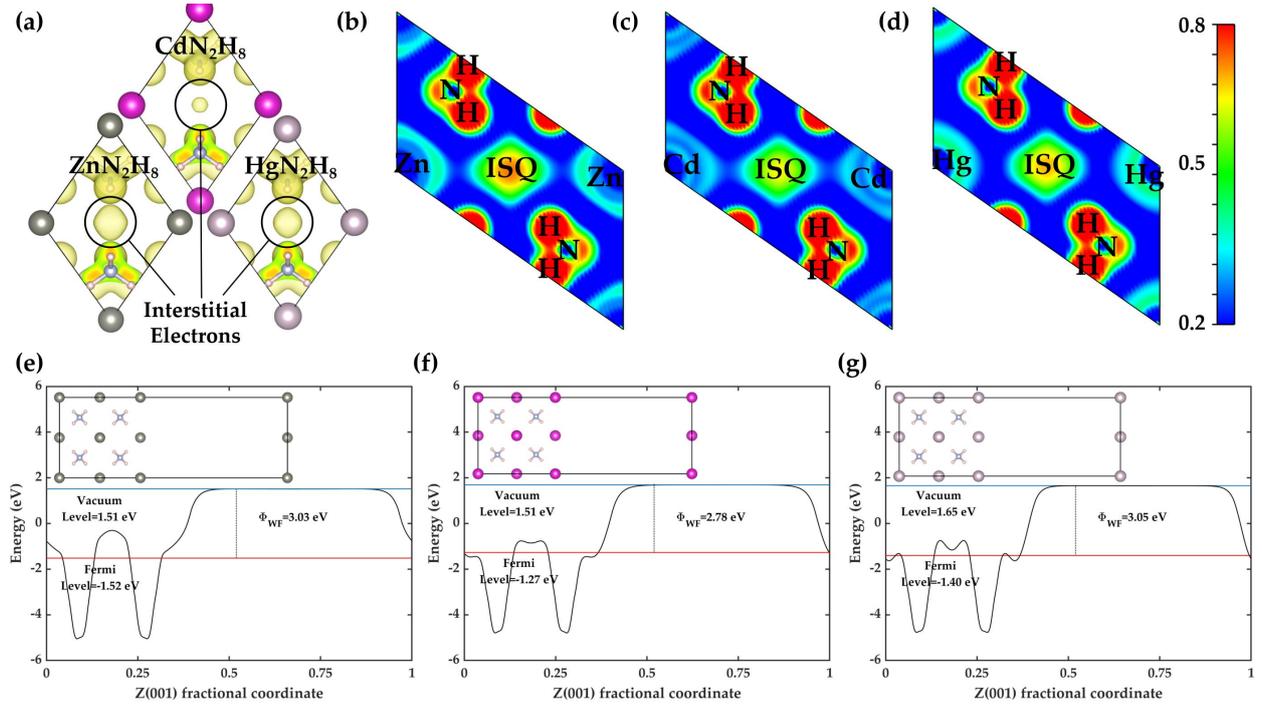

**Figure 4.** (a) Visualization of the electron localization function (isosurface=0.6). (b)-(d) Electron localization function for 2D plane (*hkl*=1,0,-1). (e)-(g) Work functions of $ZnN_2H_8$, $CdN_2H_8$, and $HgN_2H_8$ on the (001) surface, noting that (100), (010) and (001) surfaces are equivalent.

Our results also reveal that $ZnN_2H_8$, $CdN_2H_8$, and $HgN_2H_8$ can remain stable and superconducting at ambient pressure ($T_c$=0.33, 0.20, and 0.01 K), which expands the range of ambient superconducting materials. Notably, lattices composed of Ga, In, and Tl, which are very similar to the atomic structures and electronegativity of Zn, Cd, and Hg, are unstable at ambient conditions, implying that there may be exotic charge distributions in $ZnN_2H_8$, $CdN_2H_8$, and $HgN_2H_8$. The electron localization function (ELF)[32] explains for their stability with the formation of localized interstitial electrons between [$NH_4$] units (Figure 4(a)). According to Hosono et al.,[33] the ELF of ternary compounds can take the value of 0.49-0.79, so the ELF in this work is 0.6, the presence of interstitial electrons implies that they belong to electrides. Figures 4 (b)-(d) reveal the origin of the interstitial electrons. Since the 2p orbital of N can only hold up to 3 electrons, it cannot fully accept the 1s electrons from the surrounding four H atoms. The surplus electrons on [$NH_4$] enter the interstitial to reduce the total energy of the system, making the lattice more stable and acting as an interstitial quasi-atom (ISQ).[34] Figure S2 shows that $GaN_2H_8$, $InN_2H_8$, and $TlN_2H_8$ do not show interstitial electrons at 0 GPa, therefore, they cannot be stabilized at 0 GPa.

In addition, electrides can be used as catalysts,[35] and the activity of catalysts is related to the work function,[36] so the work functions ($\Phi_{WF}$) of $ZnN_2H_8$, $CdN_2H_8$, and $HgN_2H_8$ on the Z-direction (001) surface are also calculated (Figure 4(e)-(g)). Their $\Phi_{WF}$ are 3.03, 2.78, and 3.05 eV, respectively, which are much lower than Al (4.28 eV) and $Y_5Si_3$ (3.5 eV)[37] and very close to the $\Phi_{WF}$ of lithium metal (2.9 eV). In the catalytic process of nitrogen reduction reaction (NRR),[38] $N_2$ gains an electron for dissociation is a decisive step, and the lower work function means that $N_2$ can gain electrons more easily. Our results suggest that the electride $ZnN_2H_8$, $CdN_2H_8$, and $HgN_2H_8$ can be used as potential NRR catalysts.

In this work, the metallization of hydrogen was achieved by inserting ammonium ions into the fcc lattice of metals under low-pressure conditions. High throughput calculations revealed that the fcc lattice of 12 elements can have superconducting states in the range of 0-50 GPa. The superconducting transition temperatures of $AlN_2H_8$, $MgN_2H_8$, and $GaN_2H_8$ can reach up to 118.40, 105.09, and 104.39 K, respectively. Bader charge and bond length analysis reveals the physical origin of high-temperature superconductivity, and the atomization of H most likely causes the high $T_c$. The high $T_c$ is most likely caused by the atomization of H. In addition, our results identify three superconductors at 0 GPa: $ZnN_2H_8$, $CdN_2H_8$, and $HgN_2H_8$, where the interstitial electrons make them also electrides, and the presence of a relatively low work function means that they can be used as potential NRR catalysts.

The authors declare that they have no known competing interests.

The work described in this paper was supported by grants from the Research Grants Council of the Hong Kong SAR (CityU 11305618 and 11306219) and the National Natural Science Foundation of China (11874081).

Supplementary materials to this article can be found online.